# A near-perfect invisibility cloak constructed with homogeneous materials

**Wei Li, Jianguo Guan,**[*] **Zhigang Sun, Wei Wang, and Qingjie Zhang**

[1]*State Key Lab of Advanced Technology for Materials Synthesis and Processing, Wuhan University of Technology, Wuhan 430070, China*
*guanjg@whut.edu.cn*

**Abstract:** A near-perfect, non-singular cylindrical invisibility cloak with diamond cross section is achieved by a two-step coordinate transformation. A small line segment is stretched and then blown up into a diamond space, and finally the cloak consisting of four kinds and eight blocks of homogeneous transformation media is obtained. Numerical simulations confirm the well performance of the cloak. The operation bandwidth of the cloak is also investigated. Our scheme is promising to create a simple and well-performed cloak in practice.

**OCIS codes:** (160.3918) Metamaterials; (230.3205) Invisibility cloaks; (160.1190) Anisotropic optical materials; (260.2110) Electromagnetic optics.

## 1. Introduction

Based on the coordinate transformation method, Pendry et al. recently provided a theory to design a perfect invisibility cloak [1]. The device makes the cloaked region totally unreachable for electromagnetic (EM) waves and simultaneously never disturbs the EM field outside the cloak, thus an arbitrary object may be hidden in it. The perfect cloaking effect had been confirmed by the full wave numerical simulation [2-3] and further theoretical analysis [4]. An experimental attempt [5] had also been made to demonstrate the principle of such cloaks. However, practical fabrication of the full-parameter invisibility cloaks is not so far achievable, for the necessary materials are usually highly anisotropic, inhomogeneous and even singular. Therefore, great efforts have been devoted to simplifying the constitutive parameters of the materials composing the cloaks [2, 5-9]. Unfortunately, most of the simplifications are performed in a plane [10-13] and the resulted cloaks even scatter more strongly than the bare object when the EM wave is incident obliquely onto them [10]. The non-singular elliptical cloak [14] has simple constitutive parameters and good performance without simplifications. But the material parameter tensors of the elliptical cloak as well as the simplified cloaks in [10-13] are still continuously changing in the radial direction. The cloaking shell has to be cut into discrete layers when we practically fabricate the cloaks. The cloaking performance is thus inevitably reduced or even significantly decreased with respect to the perfect ones. For minimizing the degradation of the cloaking performance in practical implementation, the continuous cloaking shell must be cut into adequate layers which have different constitutive parameters. However, involving too many kinds of materials in the cloaks will again greatly complicate the fabrication. Therefore, it is still a challenge to provide a simple cloak with well cloaking performance.

Recently, Yu et al. [15] proposed and analyzed a class of invisible slab cloaks based on an "embedded optical transformation method" [16]. The slab cloak is made of homogeneous materials, which are easy to realize in practice. However, the slab seems to be infinite long according to the transformation. Thus the application of the slab cloaks is limited. Another big progress on the implementation of the cloaks is the "ground-plane cloaks" [17] which need only traditional, isotropic and readily accessible materials. The first version of the ground-plane cloak has to be embedded in a medium such as glass, and then the working background is extended to air or vacuum by a simplification [18]. However, unlike the invisibility cloaks which aim to render the cloaked object totally invisible, the ground-plane cloaks hide an object behind a conducting sheet or a mirror. Therefore, in many cases, the invisibility cloaks cannot be replaced by the ground-plane cloaks.

In this work, a cylindrical invisibility cloak with a diamond-shaped cross section is proposed based on the transformation optics. The transformation which does not expand a point as usual, but a line segment into a cloaked region thus avoids singularity in material parameters [14]. As a result, the cloak is equivalent to a small line segment in the EM fields. If the line segment is short enough, a near perfect cloak will be obtained. On the other hand, because we only compress or stretch the space in two orthogonal directions homogeneously in the transformation as we will see in the following, the resultant diamond shaped cloak consists of eight blocks of homogeneous materials and only four kinds of different materials are involved in. Thus, without any further simplifications, a near perfect cloak composed of few blocks of homogenous materials is obtained. Comparing with the simplified cloaks, the as-obtained cloak is simple and has better performance even in oblique incidence case since the cloak equivalent to a small line segment which can be far smaller that the cloaked object. Our cloak also has the potential to be a broadband one because the materials are non-singular.

## 2. Derivation of the material parameters of the diamond shaped cloak

Now let us derivate the material parameters of the diamond shaped cloak. For simplicity, we restrict ourselves in two-dimensional (2D) cases. We start from the coordinate transformation which expands a line segment to a diamond space. The transformation proceeds in two steps. Fig. 1(a)-(c) show the original Cartesian space ($x$, $y$, $z$), the transitional Cartesian space ($x'$, $y'$, $z'$), and the final transformed Cartesian space ($x''$, $y''$, $z''$), respectively. In the original space, a line segment with a length of $2a$ and midpoint at origin is parallel to the $x$ axis. The big diamond area depicted in Fig. 1a outlines the space to be transformed. The first step is aim to stretch the line segment from $2a$ to $2b$. For this purpose, a small diamond space of which the line segment is one of the diagonal lines is stretched in $x$ direction by $b/a$ times, while the space between the two diamonds is compressed homogeneously in $x$ direction. As shown in Fig. 1(b), the transformed big diamond in space ($x'$, $y'$, $z'$) is divided into eight parts I-VIII by the stretched small diamond and the coordinate axes. For each part, we can give the transformation equations. Take the two parts (region I and II) in the first quadrant as examples,

in region I:

$$x' = -\frac{(a-b)c}{(a-c)d} y + \frac{b-c}{a-c} x + \frac{(a-b)c}{a-c}, \; y' = y, \; z' = z \qquad (1)$$

in region II:

$$x' = \frac{b}{a} x, \; y' = y, \; z' = z \qquad (2)$$

The transformation equations of the other regions can also be given similarly. In fact, due to the axial symmetry of the deformation with respect to the coordinate axes, the transformation equations are symmetrical too. For example, regions IV and III are symmetrical to regions I and II, respectively. The transformation equations of them are similar to each other in form:

in region IV:

$$x' = \frac{(a-b)c}{(a-c)d} y + \frac{b-c}{a-c} x - \frac{(a-b)c}{a-c}, \; y' = y, \; z' = z \qquad (3)$$

in region III:

$$x' = \frac{b}{a} x, \; y' = y, \; z' = z \qquad (4)$$

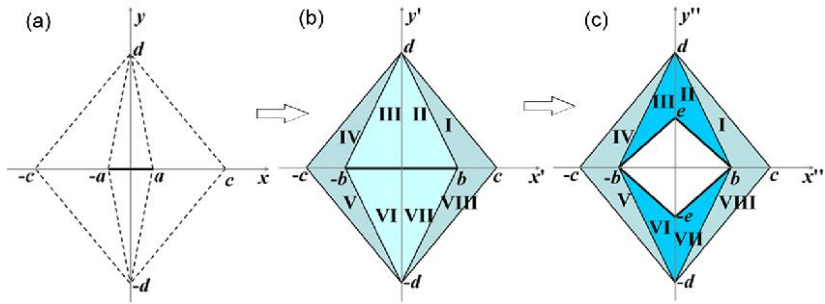

Fig. 1. The coordinate transformation for a diamond shaped cloak. (a) The original diamond shaped region with a line segment $2a$ in it. (b) The line segment and the small diamond space in (a) is stretched in $x$ direction by $b/a$ times. (c) The stretched line segment is transformed into a diamond shaped space, and the regions II, III, VI and VII are compressed. Consequently, a diamond cloak is obtained.

So far, we have stretched the given short line segment to a long one. In the second step, the stretched line segment is further expanded into a rhombic space, and the regions between the expanded rhombic space and the small diamond boundary, i.e., regions II, III, VI and VII, are compressed homogeneously in $y$ direction. Fig. 1(c) shows the final transformed space. Finally, the transformation from the transitional Cartesian space ($x'$, $y'$, $z'$) to the final transformed Cartesian space ($x''$, $y''$, $z''$) in region II is expressed as:

$$x''= x', \quad y''= \frac{d-e}{d} y' - \frac{e}{b} x' + e, \quad z''= z \tag{5}$$

The transformation in region III, VI and VII are similar to (5). In the other regions, no transformation is proceeded in the second step so we have $x''=x'$ and $y''=y'$. At the end of the transformation, a cloak containing eight blocks of material is obtained whose outer boundary and inner boundary are both diamond shaped. Combining the transformations of the two steps, the total transformation in each block is thus obtained. We still take the two regions in the first quadrant as examples,

in region I:

$$x''= -\frac{(a-b)c}{(a-c)d} y + \frac{b-c}{a-c} x + \frac{(a-b)c}{a-c}, \quad y''= y, \quad z''= z \tag{6}$$

in region II:

$$x''= \frac{b}{a} x, \quad y''= \frac{d-e}{d} y - \frac{e}{a} x + e, \quad z''= z \tag{7}$$

It should be noted that $b$, $c$, $d$ and $e$ are coordinate values related to the shape of the cloak (see Fig. 1(c)), and $a$ is half of the length of the equivalent line segment. These parameters satisfy the inequalities $c>b>a>0$ and $d>e>0$. Once we get the transformation equations, we can compute the Jacobian transformation matrix $J$ readily and then work out the relative material parameters of the cloak by (suppose the original space is free space) [3]

$$\bar{\varepsilon}_r = \bar{\mu}_r = J \cdot J^T / \det(J) \tag{8}$$

For the two regions in the first quadrant, the non-zero components of the material parameter tensors are,

in region I:

$$\begin{cases} \varepsilon_{xx} = \mu_{xx} = (A^2 + B^2)/A \\ \varepsilon_{xy} = \varepsilon_{yx} = \mu_{xy} = \mu_{yx} = B/A \\ \varepsilon_{yy} = \mu_{yy} = 1/A \\ \varepsilon_{zz} = \mu_{zz} = 1/A \end{cases} \tag{9}$$

in region II:

$$\begin{cases} \varepsilon_{xx} = \mu_{xx} = M/K \\ \varepsilon_{xy} = \mu_{xy} = MT/K \\ \varepsilon_{yy} = \mu_{yy} = K/M + MT^2/K \\ \varepsilon_{zz} = \mu_{zz} = 1/MK \end{cases} \tag{10}$$

where $A=\frac{b-c}{a-c}$, $B=\frac{(b-a)c}{(a-c)d}$, $M=b/a$, $T=e/b$ and $K=1-e/d$. The material parameters of the rest parts of the cloak can also be obtained similarly. Interestingly, the parameters in all regions are constants once the shape of the cloak and the equivalent line segment has been fixed, because no other variables are involved in the material parameters tensors in this case. Clearly, there is also no singularity in the material parameters. From Fig. 1(c), one can learn that the cloak is constructed with eight blocks of materials. Moreover, only four kinds of materials are needed indeed because the cloak is symmetrical with respect to the origin, i.e., the constitutive parameters in regions I, II, III and IV are identical to that of regions V, VI, VII and VIII, respectively. The above characteristics of the cloak highly simplify the construction of the device.

## 3. Numerical simulations and discussion

In this section, finite elements simulations are carried out to show the properties of the diamond cloak. We only consider the TM (transverse magnetic) polarization plane wave case in the simulations. In fact, similar results would be obtained in TE (transverse electric) polarization plane wave case. The plane wave is 0.15 GHz in frequency. For TM polarization, only $\varepsilon_{xx}$, $\varepsilon_{xy}$, $\varepsilon_{yx}$, $\varepsilon_{yy}$ and $\mu_{zz}$ components of the material parameters tensor are relevant. Due to the symmetry of the material parameters tensor, we have $\varepsilon_{xy}=\varepsilon_{yx}$. So we only need $\varepsilon_{xx}$, $\varepsilon_{xy}$, $\varepsilon_{yy}$ and $\mu_{zz}$ components for the simulations. The constants defining the shape of the diamond cloak are set as $b = 1$ m, $c = 2$ m, $d = 4$ m and $e = 1$ m. The half length of the equivalent line segment $a$ is chosen to be a small value of 0.15 m. A perfect electric conducting (PEC) cylinder is fitted into the cloaked region. Fig. 2(a) shows the magnetic field distribution when a TM polarized plane wave is impinging on the cloak from left to right. The wave fronts are bent around the cloaked area in the left part, and then restored their original directions by the right part, which is similar to most of the cloak deduced by the coordinate transformation method. The fields outside the cloak are almost unperturbed except tiny scattering due to the imperfection of the cloak. If the cloaked object is not a PEC cylinder which was just fitted into the cloaked region, the EM field will penetrate into the core of the device. In this case, the cloaking performance should vary with the properties such as the size, position and material parameters of the cloaked object. To avoid this, a PEC lining is usually added to the inner wall of the cloak shell.

Notice that all the materials we used in the simulation are lossless. In practice, the materials are inevitably lossy, so it is important to see what will happen when loss is introduced into the cloak. Here we simulate the case when loss tangent of permeability and permittivity are both 0.1. The result is demonstrated in Fig. 2(b), which shows strong forward scattering but hardly affects the field distribution in other directions. This result is similar to that of the previously studied cylindrical [2] and spherical cloaks [4].

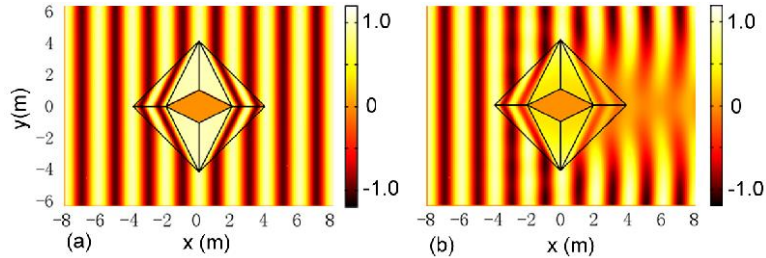

Fig. 2. Magnetic field distribution for (a) the near perfect diamond cloak without loss, and (b) the diamond cloak composed of lossy materials with loss tangent of 0.1 for both permittivity and permeability when a TM plane wave is incident from left to right.

Since the diamond cloak equivalents to a line segment instead of a point, it will have variant performances when the EM wave is incident from different directions. Specifically, when the incident wave is perpendicular to the equivalent line segment, the cloak will show the strongest scattering. Numerical simulation has also been carried out to examine the performance of the cloak in this case. Fig. 3(a) gives the magnetic field distribution of the result which shows a tiny scattering. However, if the equivalent line segment is too long, strong scattering will be induced by the cloak along the direction perpendicular to the line segment. Fig. 3(b) shows the case when $a = 0.5$ m. To further qualify these results, the total scattering cross sections versus $a$ are plotted in Fig. 4(a) when the incident wave is perpendicular, parallel and at a 45 degree angle to the $+x$ axis respectively, From Fig. 4(a) we can learn that the longer the equivalent line segment, the stronger scattering the cloak will induce. With the increasing of the wave incident angle, the total scattering also increases quickly. However, the scattering can be kept at a very low level so long as the equivalent line segment is short.

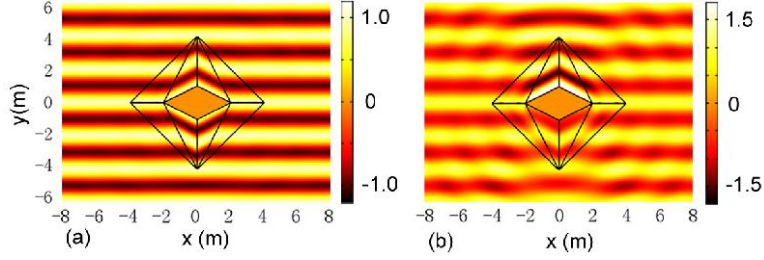

Fig. 3. Magnetic field distribution when a TM plane wave is impinging on the cloak with an effective line segment length $2a$ = (a) 2×0.15 m and (b) 2×0.5 m from top to bottom.

As discussed above, the length of the equivalent line segment decides the cloaking performance when the incident wave is not parallel to the line, but it does not complicate the construction of the cloak, i.e., the cloak is still made up of a few blocks of homogenous materials. In this sense, we can achieve an uncomplicated cloak arbitrarily near the perfect one by choosing a small value of $a$. However, it should be noticed that an extreme small value of $a$ may also lead to extreme material parameters, which are hard to be realized, according to the material parameters equations such as Eq. (9) and (10). With regard to the other parameters, namely, $b$, $c$, $d$ and $e$, they are irrelevant to the performance of the cloak in principle. However, they are closely related to the material parameters of the cloak. We can adjust them to set the material parameters within a desirable range at the price of increasing the thickness of the cloak. Finally, considering fabricating the cloak, it is preferable to denote the material parameters in their eigenbasis, where the permittivity and permeability tensors are diagonal. Fortunately, due to the symmetry of the tensors, one can always do this. For example, the material parameters of region I (see Eq. (9)) can be express as follow:

$$\bar{\varepsilon}_r = \bar{\mu}_r = \begin{pmatrix} (1+A^2+B^2+|A^2+B^2-1|)/2A & 0 & 0 \\ 0 & (1+A^2+B^2-|A^2+B^2-1|)/2A & 0 \\ 0 & 0 & 1/A \end{pmatrix} \quad (11)$$

An experimental implementation of the device may be achieved by anisotropic metamaterials such as split-ring resonators (SRRs) [5], canonical spirals [19] at microwave frequencies, and photonic crystals at optical frequencies. However, metamaterials are not always needed. By using the effective medium theory (EMT) one can also construct the anisotropic devices with layered isotropic materials [20]. A related experimental work is now being considered.

The bandwidth has been a big issue for traditional transformation optics based cloaks, which work only at a single frequency because of the singularities in the material parameters of them. The as-obtained diamond-shaped cloak is non-singular, thus it is expected to operate at broad bandwidth. Here we investigate the cloaking performance of the diamond-shaped cloak in the range of 0.1-0.2 GHz when a plane wave is incident from top to bottom. Note that the device is dispersive like most of the reported cloaks. To introduce the dispersion into the cloak, we assume that the material parameters of the cloak follow the standard Lorentz model:

$$\eta_j(\omega) = 1 + [\eta_j(\omega_0) - 1] f_j(\omega) / \mathrm{Re}[f_j(\omega_0)] \qquad (12)$$

where $\eta_j$ represents $j$ component of permittivity or permeability tensor, and $f_j(\omega) = \omega_p^2/(\omega_{tj}^2 - \omega^2 - i\omega\gamma)$ are the Lorentzian dispersive functions, where $\gamma$ is the damping factor, $\omega_p$ is the plasma frequency, $\omega_{tj}$ is the atom resonated frequency. We set $\omega_0 = 2\pi \times 0.15$ GHz, $\omega_p = 4\omega_0$, $\gamma = 0.01\omega_0$, $\omega_{tj} = 0.6\omega_0$ when $\eta_j < 1$ and $\omega_{tj} = 1.4\omega_0$ when $\eta_j \geq 1$. Fig. 4(b) shows that the normalized total scattering cross section varies with the frequency. At frequency of 0.15 GHz, the total scattering is quite small. However, as the EM wave frequency deviates from that point, the total scattering increases drastically because of the dispersion and the losses brought by the dispersion. In spite of that, the cloak is found to perform well in a certain range of frequencies. In the range of 0.13-0.17 GHz, the total scattering cross section can be reduced 50% at least. The result suggests that the diamond-shaped cloak do work at multiple frequencies other than a single frequency like the singular ones, though the bandwidth is limited due to the dispersion and losses of the cloak. It is imaginable that different dispersion parameters or different materials as well as the parameters of the cloak such as the length of the equivalent line segment may lead to different bandwidth, which we will discuss elsewhere.

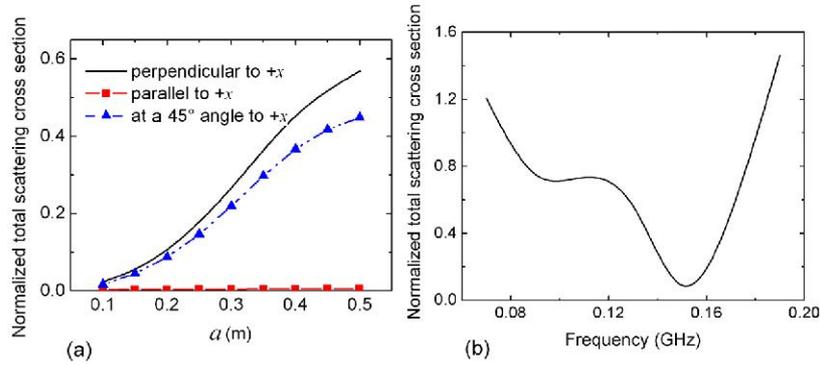

Fig. 4. Normalized total scattering cross section (normalized to the diamond PEC cylinder) as a function of (a) the half length of the equivalent line segment $a$ at 0.15 GHz when the incident wave is perpendicular (black line), parallel (blue triangle) and at a 45 degree angle to the $+x$ axis (red rectangular) respectively for the non-dispersive diamond-shaped cloak, and (b) frequencies when $a=0.15$ m and the incident wave is perpendicular to $+x$ axis for the dispersive diamond-shaped cloak.

## 4. Summary

In summary, we have presented a near-perfect diamond-shaped cloak based on a two-step coordinate transformation. Since the transformation expands a line segment instead of a point into the cloaked region, singularities are avoided in the material parameters of the device. Moreover, due to homogeneously stretching and compressing at two orthogonal directions during the transformation, the resulted transformation media are homogeneous. Numerical simulations reveal that the cloak has a good cloaking performance for all the incident directions of plane waves and can work at multiple frequencies. The cloak is simply constructed by only four kinds of non-singular homogeneous materials, which make the device easy to realize in practice.

**Acknowledgements**

This work was supported in part by the National High-Technology Research and Development Program of China under Grant No. 2006AA03A209, in part by the National Defense Basic Research project under Grant No. D1420061057, in part by the Young Teacher Grant from Fok Ying Tung Education Foundation under Grant No. 101049 and in part by the Ministry of education of China under Grant No. PCSIRT0644.